\documentclass[pre,twocolumn,superscriptaddress,aps]{revtex4-1}
\usepackage{empheq}
\usepackage{amsmath}
\usepackage{tabularx}
\usepackage{dsfont}
\usepackage{graphicx}
\usepackage[hidelinks]{hyperref}
\usepackage{stackrel}
\usepackage{amssymb,latexsym,mathrsfs}
\usepackage{amsfonts}
\usepackage{srcltx}
\usepackage{color}
\usepackage{cancel}
\usepackage[normalem]{ulem}
\usepackage{xr}
\usepackage{xspace}
\usepackage{placeins}
\usepackage{stix}

\newcommand{\lmin}{L_\text{min}}
\newcommand{\la}{\langle}
\newcommand{\ra}{\rangle}
\newcolumntype{Y}{>{\centering\arraybackslash}X}
\newcolumntype{L}[1]{>{\raggedright\let\newline\\\arraybackslash\hspace{0pt}}m{#1}}
\newcolumntype{C}[1]{>{\centering\let\newline\\\arraybackslash\hspace{0pt}}m{#1}}
\newcolumntype{R}[1]{>{\raggedleft\let\newline\\\arraybackslash\hspace{0pt}}m{#1}}

\newcommand{\om}{$\omega$\xspace} 
\newcommand{\redchi}{$\chi^2$/d.o.f\xspace}

\begin{document}

\title{Universality of continuous phase transitions on random Voronoi graphs}
\author{Manuel Schrauth}
\author{Jefferson S. E. Portela}
\affiliation{Institute of Theoretical Physics and Astrophysics,	University of W\"urzburg, 97074 W\"urzburg, Germany}

\begin{abstract}
	The Voronoi construction is ubiquitous across the natural sciences and engineering. In statistical mechanics, though, critical phenomena have so far been only investigated on the Delaunay triangulation, the dual of a Voronoi graph. In this paper we set to fill this gap by studying the two most prominent systems of classical statistical mechanics, the equilibrium spin-1/2 Ising model and the non-equilibrium contact process, on two-dimensional random Voronoi graphs. Particular motivation comes from the fact that these graphs have vertices of constant coordination number, making it possible to isolate topological effects of quenched disorder from node-intrinsic coordination number disorder. Using large-scale numerical simulations and finite-size-scaling techniques, we are able to demonstrate that both systems belong to their respective clean universality classes. Therefore, quenched disorder introduced by the randomness of the lattice is irrelevant and does not influence the character of the phase transitions. We report the critical points of both models to considerable precision and, for the Ising model, also the first correction-to-scaling exponent.
\end{abstract}

\maketitle

\section{Introduction}

\begin{figure}[b]
	\centering
	\includegraphics[width=0.91\linewidth]{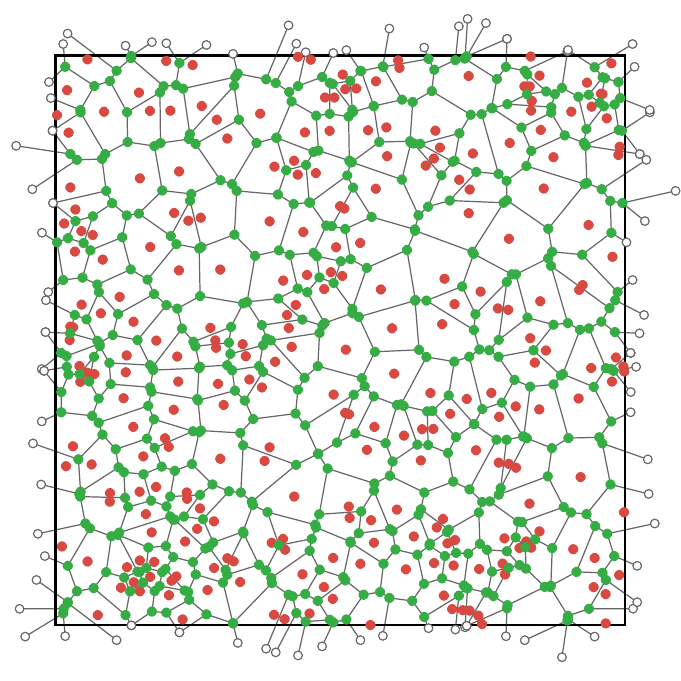}
	\vspace*{-2mm}
	\caption{[Color online] Voronoi graph (green) with periodic boundaries, generated from a set of a Poissonian random points (red).}
	\label{fig:lattice}		
\end{figure}

Many location optimization problems can be approached through area-of-influence considerations. A simple example is that of several fire stations distributed over a large city. Rather naturally, the area of responsibility attributed to a particular fire station should include those buildings which lie closer to it than to any other station. The resulting tessellation of the city map defines the so-called Voronoi graph (VG). In Fig.~\ref{fig:lattice}, the red dots would denote fire stations, with the corresponding VG being depicted in green. Due to its conceptual simplicity, Voronoi constructions can be found in a large number of applications spanning all fields of physical sciences, including climate modeling~\cite{ringler2008,ju2011}, crystal structure~\cite{wigner1933}, cosmology~\cite{weygaert1989,ramella2001}, microbiology \cite{poupon2004}, and growth processes~\cite{william1939}, as well as optimization problems~\cite{okabe1997}, game theory~\cite{ahn2004,cheong2004,durr2007}, artificial intelligence \cite{forbus2002,gavrilova2008} and, recently, also in the field of machine learning~\cite{ward2017}, among others. Moreover, numerous generalizations have been defined, such as weighted graphs, and Voronoi graphs on spherical and general curved surfaces,~\cite{na2002,boissonnat2006}, as well as for fuzzy point sets~\cite{jooyandeh2009} or metrics other then Euclidean~\footnote{In a more realistic fire station example one might consider using the Manhattan distance rather than the Euclidean one}.

In this paper, the VG is used as a discrete spatial lattice on top of which the physical system is evolved. During the evolution the lattice remains static, i.e., we consider \emph{quenched} disorder~\cite{binder1986}. As the graph is constructed from a randomly distributed set of points it is said to present \emph{topological} randomness, as opposed to conventional disorder, as e.g., in the case of randomly diluted regular structures.

Quenched disorder in general constitutes a relevant perturbation on the phase transition, i.e., it may change the universal properties of the model~\cite{nishimori2010}. The first major result on disorder relevance was brought forward by Harris in 1974~\cite{harris1974}, who argued that disorder is relevant when $d\nu<2$, where $d$ is the dimension of the system and $\nu$ denotes the correlation length exponent. Using Harris' criterion, numerical results on 2D diluted regular lattices can be entirely explained, including the rise of non-conventional activated scaling and strong Griffiths effects for the contact process (CP)~\cite{vojta2005,vojta2009,wada2017}, as well as ambiguities related to strong logarithmic corrections for the Ising model~\cite{kim1994b,ziegler1994,kuehn1994,dotsenko1981,jug1983,shankar1987,martins2007,fytas2013,zhu2015}, which represents the marginal case, as in two dimensions $\nu=1$. Despite its success in describing the effects of uncorrelated disorder, the Harris criterion fails to explain certain results for the Ising model and CP on two-dimensional Delaunay triangulations~\cite{janke1993,janke1994,lima2000,janke2002,oliveira2008,oliveira2016}, where both systems retain their clean universal properties. In order to explain these results, a generalization has been proposed~\cite{barghathi2014}, which attributes the non-relevance to strong spatial anti-correlations in the coordination numbers of the lattice nodes. Also for this new criterion, however, violations have been found~\cite{schrauth2018b}. Moreover, as the argument specifically relies on fluctuations in the local coordination numbers, this criterion is silent about random structures with \emph{constant} coordination number, such as the VG, where each site has exactly three neighbors (compare Fig.~\ref{fig:lattice}).

Therefore, finding a complete criterion for the influence of topological disorder on continuous phase transitions still remains an unsolved puzzle, to which the study of critical phenomena on constant coordination lattices can provide valuable new pieces. One such puzzle piece is given in the current work, where numerical simulations of the Ising model and the CP on two-dimensional VGs show clean universal behavior for both systems, revealing the perturbations by quenched topological randomness to be irrelevant for these systems.

The paper is organized as follows. In Sec.~\ref{sec:VoronoiGraph} we review the construction rules of the VG. In Sec.~\ref{sec:IsingModel} and \ref{sec:DirectedPercolation} we present and analyze numerical simulations of the Ising model and the CP, respectively. Finally, in Sec~.\ref{sec:Conclusion} we present our concluding remarks.

\section{Voronoi Graph}
\label{sec:VoronoiGraph}

The Voronoi diagram is a partition of the plane into cells, generated by a set of points $P= \{p_1, p_2, \dotsc,p_N\}$ such that for each cell corresponding to the point $p_i$, every point $q$ in that cell is closer to $p_i$ than to any other point $p_j$, i.e., $d(q,p_i) < d(q,p_j)\,\,\forall p_j\neq p_i$. It is the geometric dual of the Delaunay triangulation, which can be constructed by connecting points corresponding to adjacent Voronoi cells.
The VG is defined by taking as sites the corners of the cells and, as edges, the boundaries between the cells. Therefore the new set of points $P'= {p'_1, p'_2, \dotsc,p'_{2N}}$ is twice as large as the original set that defined the cells. This can be easily seen from the Euler characteristic of a finite graph, which is defined as $\chi = N-E+F$, where $N$, $E$ and $F$ are the number of vertices, edges and faces. For the periodic boundary conditions used here $\chi=0$ holds. In a Delaunay triangulation, $E=3N$ as the average coordination number is exactly $q=6$ and any edge is shared by two triangles. Therefore we end up with $F=2N$ faces in the triangulation and hence $2N$ points in the VG due to the duality property. The location of the VG sites is given by the center of the circumcircle of the corresponding triangle in the Delaunay triangulation. A sample of a periodic VG is shown in Fig.~\ref{fig:lattice}. It can be easily seen that all cells have convex shape and that every site has exactly three neighbors. This latter feature, the absence of coordination number fluctuations, constitutes a major motivation of this study. We build VGs from Delaunay triangulations, and for constructing the initial Delaunay triangulations we use the CGAL library~\cite{kruithof2017}.

\section{Ising Model}
\label{sec:IsingModel}

The classical Ising model~\cite{ising1925} is defined by the Hamiltonian
\begin{align}
	\mathcal{H} = -\sum \limits_{ \langle i,j \rangle}J_{ij} s_i s_j +\sum \limits_{i}h_is_i , \quad\quad s_i = \pm 1
	\label{eq:Hamiltonian}
\end{align}
where $s_i$ are discrete spins on the lattice sites, $J_{ij}$ denotes the coupling between nearest neighbors $\langle i,j \rangle$ and $h_i$ is the external field at site $s_i$. For equilibrium lattice models, quenched disorder can be introduced in a variety of ways. For fixed ferromagnetic coupling $J_{ij} = J >0 $ and random external field, the system is called random field Ising model (RFIM) and has been investigated thoroughly over the last decades~\cite{nattermann1998}. In contrast, for vanishing external field but randomly distributed (anti)ferromagnetic bonds, the system shows the behavior of a spin glass~\cite[and references therein]{binder1986}. In the present study, quenched disorder is introduced as topological randomness encoded in the implicit connectivity of the Voronoi graph. We therefore fix all couplings to $J_{ij}=1$ at vanishing external field.

For the simulation of the Ising model at criticality state-of-the-art importance-sampling Monte Carlo methods are employed, using cluster, as well as local update algorithms. In particular, we use the algorithm proposed by Wolff \cite{wolff1989}, which significantly reduces the critical slowing down and is furthermore straightforwardly applicable to disordered lattices. Although the cluster updates preserve ergodicity, we add local Metropolis updates~\cite{metropolis1953} in order to make sure that the short-wavelength modes are properly thermalized.

In the study of disordered systems, it is necessary to average physical observables over many different, independent disorder realizations of the system, also called \emph{replicas}. The quenched averages over $N_r$ replicas are performed at the level of (extensive) observables, rather than at the level of the partition function~\cite{binder1986}. Denoting quenched averages as
\begin{align}
	\label{eq:disorder_average}
	[\mathcal{O}]_\text{avg} \equiv \frac{1}{N_r} \sum \limits_{i=1}^{N_r} \mathcal{O}_i
\end{align}
and thermal averages as $\langle ... \rangle $,
we define magnetization, energy and susceptibility as
\begin{subequations}
	\begin{align}
	M &= [ \la |m| \ra ]_\text{avg}, \\
	E &= [ \la e \ra ]_\text{avg}, \\
	\label{eq:obs_chi}
	\chi &= N \beta [ \la m^2 \ra - \la |m| \ra^2  ]_\text{avg},
	\end{align}
\end{subequations}
respectively, where $m$ and $e$ denote the magnetization and energy per site. Furthermore, the two-point finite-size correlation function is given by
\begin{align}
	\xi =\dfrac{1}{2\sin(k_\text{min}/2)} \sqrt{\dfrac{[\large\langle\large|\mathcal{F}^2(\mathbf{0})\large|\large\rangle]_\text{avg}}{[\large\langle|\mathcal{F}^2(\mathbf{k}_\text{min})\large|\large\rangle]_\text{avg}}-1},
\end{align}
with the Fourier transform of the magnetization being defined by
\begin{align}
	\mathcal{F}(\mathbf{\mathbf{k}}) = \sum_{j} s_j\exp(i\mathbf{kx}_j),
\end{align}
where $\mathbf{x}_j$ denotes the spatial coordinate of site $j$ and $  \mathbf{k}_\text{min}=(2\pi/L,0) $ represents the smallest non-zero wave vector in the finite lattice. The time series of the quantities $m$, $e$ and $ \mathcal{F}^2(\mathbf{k}_\text{min}) $ is recorded during the Monte Carlo run, which allows to compute all relevant observables subsequently. Finally, the fourth- and sixth-order magnetic cumulant, also called Binder ratios~\cite{binder1981}, are given by
\begin{align}
	U_4 = \Bigg[ 1 - \frac{ \la m^4 \ra }{ 3 \la m^2 \ra^2}  \Bigg]_\text{avg},\quad U_6 = \Bigg[ \frac{ \la m^6 \ra }{ \la m^2 \ra^3}  \Bigg]_\text{avg}.
\end{align}
These quantities together with the correlation length in unit of the linear system size share the property of being invariant under RG transformation, i.e., they tend towards well-defined fixed point values as the system approaches the critical point. As a first step in our analysis we determine these values for $R\equiv\{\xi/L,U_4,U_6\}$ as well as the eigenvalue \om of the first irrelevant RG operator using the quotient method~\cite{ballesteros1996,ballesteros1997b,ballesteros1998a}. This finite-size scaling (FSS) technique allows for great statistical accuracy and does not require a precise knowledge of the critical temperature. Specifically, the scaling functions $R$ are evaluated at crossing points of $ \xi/L $ curves, where they are expected to scale as
\begin{align}
	R|_{Q_\xi=s}=R^*+a_R L^{-\omega},
	\label{eq:Rstar_fit}
\end{align}
neglecting additional sub-leading correction terms. Here, \om is the leading correction-to-scaling exponent, $R^*$ denotes the value of the RG-invariant scaling functions at the fixed point and the amplitudes $a_R$ of the leading corrections depend on the respective scaling functions $R$. Furthermore, in Eq.~\eqref{eq:Rstar_fit} the quotients are defined as $Q_O = O(sL,T)/O(L,T)$, which means that the observable $O$ -- in our case $\xi$ -- is measured at a temperature for which the correlation length in units of the linear lattice size is the same for the pair $ (L,sL) $. We fix $ s=2 $. 

Using histogram reweighting techniques~\cite{ferrenberg1988,ferrenberg1989} we determine the crossing points of $ \xi/L $ precisely. The reweighting procedure is performed for every disorder replica individually and the curves are averaged afterwards. Up to $ 10^5 $ disorder realizations are used for the smallest lattices and at least 4000 replicas for the largest ones. Every replica is initially prepared in a cold spin configuration and is thermalized using 1000 Elementary Monte-Carlo step (EMCS). We checked for a proper thermalization by also performing simulations starting from a hot configuration, which gives identical results within numerical precision. In our update procedure, one EMCS consists of a full Metropolis lattice sweep and several single-cluster updates. Since the average cluster size $\langle|C|\rangle$ in a $d$-dimensional system at criticality scales as $ L^{d-\gamma/\nu} $, we increase the number of cluster updates with lattice size according to $L^{0.25}$ in order to keep the fraction of flipped spins approximately independent of the lattice size~\cite{janke2002}. A detailed list of replica configurations and cluster steps can be found in Tab.~\ref{tab:simulation_settings}. The simulations reported in this section took about half a million CPU hours on an Intel Xeon E5-2697 v3 processor.

\begin{table}[b]
	\small
	\centering
	\caption{Results of the simultaneous fits according to Eq.~\eqref{eq:Rstar_fit}.}
	\begin{tabularx}{\columnwidth}{C{7mm}C{16mm}C{16mm}C{16mm}C{13mm}Y}
		\hline
		\hline
		$\lmin$ & $ (\xi/L)^* $ & $ U_4^*$ & $U_6^*$ & \om & $\chi^2$/d.o.f \\
		\hline
		16 & 0.9078(2) & 0.61067(2) & 1.4563(2) & 1.36(2) & 12.6 \\
		18 & 0.9070(2) & 0.61066(2) & 1.4564(2) & 1.43(2) & 7.2 \\
		20 & 0.9066(3) & 0.61066(2) & 1.4564(2) & 1.47(3) & 5.8 \\
		24 & 0.9062(3) & 0.61065(2) & 1.4564(2) & 1.53(4) & 5.0 \\
		32 & 0.9058(3) & 0.61065(3) & 1.4563(2) & 1.59(7) & 4.7 \\
		40 & 0.9060(5) & 0.61067(4) & 1.4561(3) & 1.50(12) & 4.1 \\
		48 & 0.9063(7) & 0.61069(5) & 1.4559(4) & 1.37(18) & 4.5 \\
		\hline	
		\hline
	\end{tabularx} 
	\label{tab:fit_results_Rstar_joint}
\end{table}

In order to obtain the fixed point phenomenological couplings and the correction exponent, we perform simultaneous fits of Eq.~\eqref{eq:Rstar_fit} for all three couplings $\{\xi/L,U_4,U_6\}$ with joint \om and for different $\lmin$, i.e., discarding the smallest lattices in the fits. As the quotients $Q_O$ are naturally correlated in pairs $ (L,2L) $, we implement a fitting procedure that optimizes a generalized $ \chi^2 $, including the full self-covariance information, as proposed in Ref.~\cite{gordillo2007}. Uncertainties for the fit parameters are obtained by bootstrap resampling methods~\cite{efron1994}\footnote{We construct 250 bootstrap samples of the full data set by averaging the observables over $N_r(L)$ randomly drawn disorder replicas rather than performing a simple average where every replica is considered exactly once. For every of those bootstrap samples the fits are performed, resulting in 250 estimates for the fit variables. Averages and standard deviations of these estimates are reported as the final results.}. The results are shown in Tab.~\ref{tab:fit_results_Rstar_joint}. As $\lmin$ is increased, the fit results show slight systematic trends, caused by higher-order corrections. Above $\lmin\approx 24$ the values saturate and the \redchi of the fit does not improve further. 
We therefore use, as our final estimates, the averages for $ \lmin=24,32,40 $ and adopt a rather conservative uncertainty which includes the fluctuations among the single estimates as well as their individual uncertainties. This yields, as our final results
\begin{align}
\omega = 1.54(16)
\label{eq:ResultOmega}
\end{align}
and 
\begin{subequations}
	\begin{align}
		(\xi/L)^* &= 0.9060(5)\\
		U_4^* &= 0.61066(3)\\
		U_6^* &= 1.4563(5).
	\end{align}
\end{subequations}
Comparing our estimates for the critical couplings with reference values of the Ising model on a regular square lattice, which are known exactly, up to small uncertainties from numerical integration, $ (\xi/L)^* = 0.90505\dotso $, $ U_4^* = 0.61069\dotso $, $U_6^* = 1.45565\dotso$~\cite{kamieniarz1993,salas2000} we find that even though these quantities are only considered universal in a limited sense (they weakly depend on certain geometrical characteristics of the system \cite{selke2005,selke2009,malakis2014}) they compare remarkably well, giving a first indication that the Ising model on a VG stays in the universality class of the clean model. Also our result for the correction exponent is noticeably smaller than the reference value on a square lattice, $\omega=2$~\cite{blote1988}, though not particularly small in absolute numbers, which explains why corrections to scaling turn out to be relatively weak in the scaling collapses described below.

\begin{figure}
	\centering
	\includegraphics[width=\linewidth]{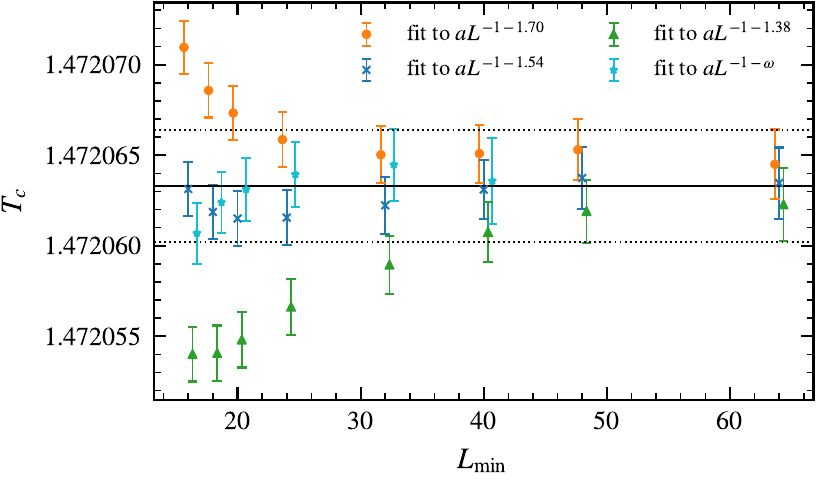}
	\caption{Estimates for $T_c$ from fits to Eq.~\eqref{eq:TcritFit}. Some results are slightly shifted along the x-axis to make them visible. The horizontal lines correspond to the final estimate $T_c = 1.4720633(31)$.}
	\label{fig:TcLmin}		
\end{figure}

Essential for computing scaling collapses is a precise knowledge of the location of the critical point, which depends on the details of the lattice structure and is therefore in general not known in advance. In the framework of quotient-FSS the critical temperature can be obtained using infinite-volume extrapolations, as the crossing points are expected to scale according to~\cite{binder1981}
\begin{align}
	T|_{Q_\xi=s}=T_c+a L^{-\omega-1/\nu},
	\label{eq:TcritFit}
\end{align}
where higher-order terms have been neglected as in Eq.~\eqref{eq:Rstar_fit}, and we adopt the clean exponent $\nu=1$. We perform four series of fits, where in the first three the correction exponent is fixed to our previous estimate \eqref{eq:ResultOmega}, plus and minus its uncertainty, i.e., $\omega=1.38,1.54,1.70$. In the last series of fits \om is treated as a free parameter. The results are displayed in Fig.~\ref{fig:TcLmin} and listed in detail in the Supplementary Material, including \redchi values. It can be seen that for $\lmin\gtrsim40$ all four fits are compatible within their error bars. As our final estimate we take the average of the fixed-\om fits for $\lmin=64$, obtaining
\begin{align}
	T_c = 1.4720633(31).
	\label{eq:Tcrit}
\end{align}

\begin{table}[b]
	\small
	\centering
	\caption{Number of disorder replicas and cluster updates per EMCS for the Ising simulations at criticality. All systems were simulated at $T=1.47205$.}
	\begin{tabularx}{\columnwidth}{L{1mm}L{3cm}L{2.9cm}L{3cm}}
		\hline
		\hline
		&$L$ & $N_r$ & $n_\mathrm{Wolff}$ \\
		\hline
		&16,\,18,\,20,\,24 & 100\,000 & 10 \\
		&32,\,36,\,40,\,48 & 100\,000 & 12 \\
		&64,\,80,\,96 & 100\,000 & 15 \\
		&128,\,192,\,256 & 35\,000 & 17 \\
		&384,\,512 & 15\,000 & 23 \\
		&768 & 5000 & 27 \\
		&1024 & 4000 & 29 \\
		\hline
		\hline
	\end{tabularx} 
	\label{tab:simulation_settings}
\end{table}

In the next step of the analysis we simulate the Ising model on Voronoi graphs of size $L=24,32,\dotsc,384$ for several temperatures in the vicinity of the critical point, using at least $1000$ disorder replicas for every lattice size and temperature. Similar to the precision simulations directly at criticality reported above, we start from cold configurations and perform 2500 measurements after a thermalization time of 500 EMCS. The number of cluster updates per EMCS is reduced by about a factor of five with respect to the values reported in Tab.~\ref{tab:simulation_settings}.

\begin{figure}
	\centering
	\includegraphics[width=\linewidth]{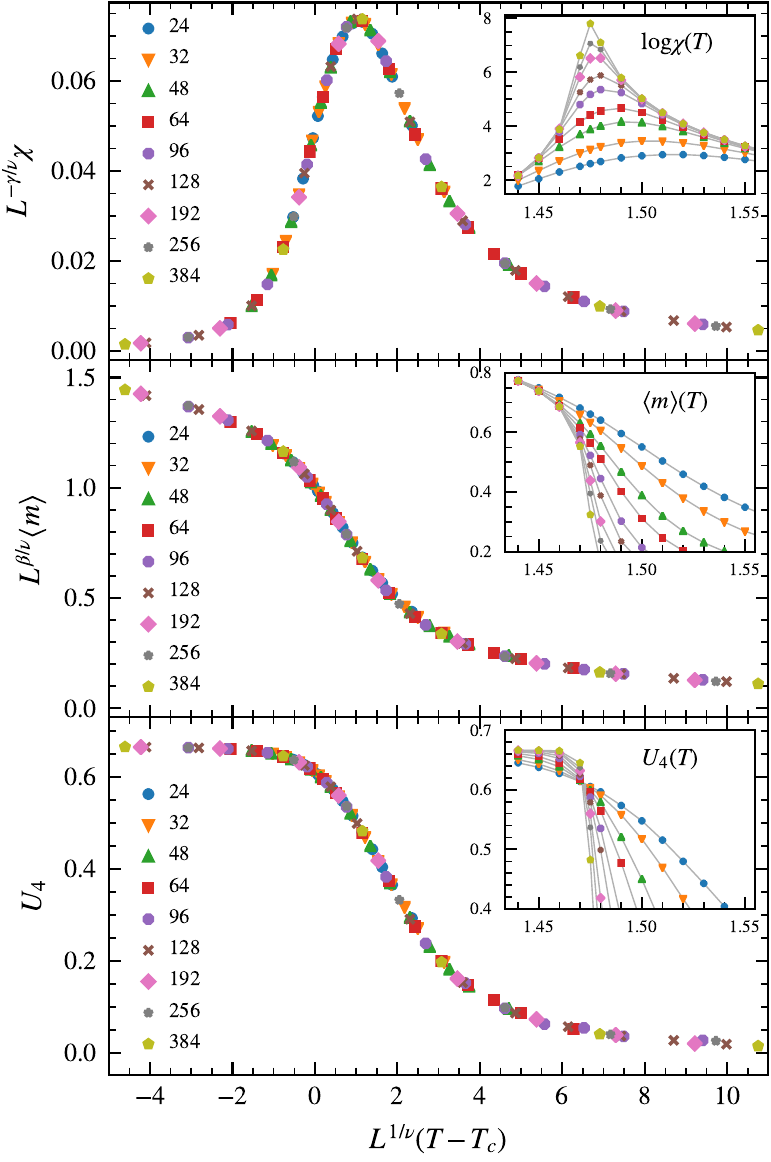}
	\caption{Scaling collapses according to Eq.~\eqref{eq:scaling_relations}. The insets show the non-rescaled observables. The gray lines are only guides to the eye.}
	\label{fig:collapse}		
\end{figure}

Finite-size scaling theory predicts that magnetization, susceptibility and Binder ratio scale according to 
\begin{subequations}
	\label{eq:scaling_relations}
	\begin{align}
	\label{eq:scaling_m}
	[ \la m \ra ]_\text{avg} &= L^{-\beta/\nu} f_m(x)(1+\dotsb), \\
	\label{eq:scaling_chi}
	\chi &= L^{\gamma/\nu} f_\chi(x)( 1 + \dotsb ),\\
	U_4 &= f_{U_4}(x)( 1 + \dotsb ),
	\end{align}
\end{subequations}
where $\beta$, $\gamma$ and $\nu$ are critical exponents of the model and the universal scaling functions $f$ have the argument
\begin{align}
x \equiv (T - T_c)L^{1/\nu}.
\end{align}
	
These equations describe the scaling behavior to first order. Corrections of higher order are expected to become irrelevant for large system sizes. In Fig.~\ref{fig:collapse} we show the scaling collapse plots, fixing all critical exponents to their exactly known values ($\nu=1$, $\beta=1/8$, $ \gamma=7/4 $) and $T_c$ to the estimate~\eqref{eq:Tcrit}. As can be seen, a flawless collapses for all three scaling functions is obtained even for small lattices, which shows that the Ising model on a 2D random Voronoi graphs belongs to the universality class of the clean 2D Ising model.

\section{Directed Percolation}
\label{sec:DirectedPercolation}
	
The most prominent family of non-equilibrium phase transitions arguably is the so-called directed percolation (DP) universality class. According to the Janssen-Grassberger DP conjecture, any system featuring a fluctuating phase and a unique absorbing state with scalar order parameter, no additional symmetries and only local interactions falls into this class~\cite{janssen1981,grassberger1982}. On a lattice, the DP universality class is realized by certain reaction-diffusion schemes, such as the contact process (CP). In this model every site can be in either of two states, active or inactive. In the epidemic language, often used in this context, one refers to infected and recovered states. The system is evolved as a Markov process, where in each time step, one random active site is picked. With probability $p$ it infects a random neighbor whereas with probability $1-p$ the particle spontaneously recovers and is removed from the active cluster. Time is then incremented by $1/N_a$, where $N_a$ denotes the size of the active cluster before the update. As soon as the system enters the so-called absorbing state where every site is inactive, the dynamics terminates. 
	
\begin{figure}
	\centering
	\includegraphics[width=\linewidth]{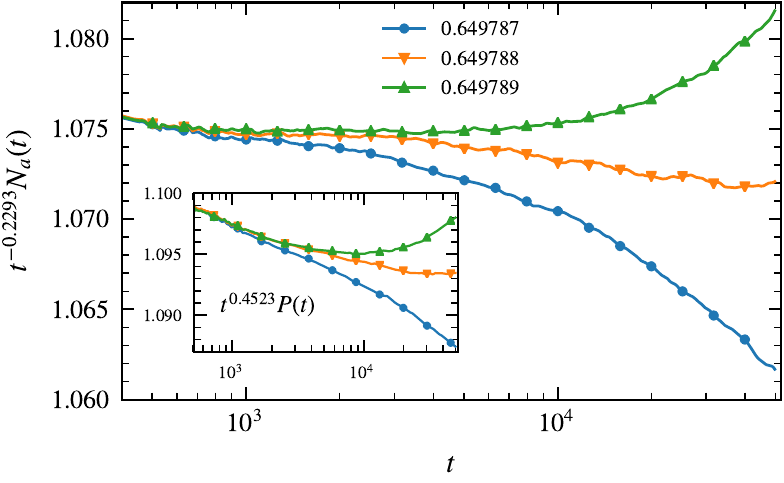}
	\caption{Size of the active cluster for different probabilities $p$ in seed simulations of the CP close to the critical point, rescaled by the expected universal scaling law, $N_a(t)\sim t^\theta$. The inset shows the survival probability of the process, rescaled according to $P(t)\sim t^{-\delta}$. For the critical exponents, reference values from Ref.~\cite{dickman1999} are used.}
	\label{fig:contact-pc}		
\end{figure}

In order to show that the CP on a random VG shows clean universal behavior we conduct numerical simulations as described in Ref.~\cite{schrauth2018b}, in total amounting to approximately 300\,000 CPU hours on an Intel Xeon E5-2697 v3 processor. As a first step, from seed simulations we determine the critical point, by rescaling the cluster size and survival probability according to their expected power law behavior, $N_a(t)\sim t^\theta $ and $ P(t)\sim t^{-\delta} $, where $\theta$ and $\delta$ denote critical exponents. In total, we use $ 10^5 $ independent disorder realizations of linear size $ L=2048 $ with periodic boundary conditions and performed $ 10^4 $ seed runs on each of them. Using reference values from Ref.~\cite{dickman1999}, $\theta=0.2293(4)$ and $ \delta=0.4523(10) $, we obtain the critical probability
\begin{align}
	p_c=0.649788(1) 
\end{align}
by determining the asymptotically constant curve, as shown in Fig.~\ref{fig:contact-pc}. The uncertainty is determined from curves which noticeably bend away from horizontal behavior. Note that using larger lattices and longer simulation times would not significantly increase the precision of this estimate, as the analysis is limited by the uncertainties of the reference values.

\begin{figure}
	\centering
	\includegraphics[width=\linewidth]{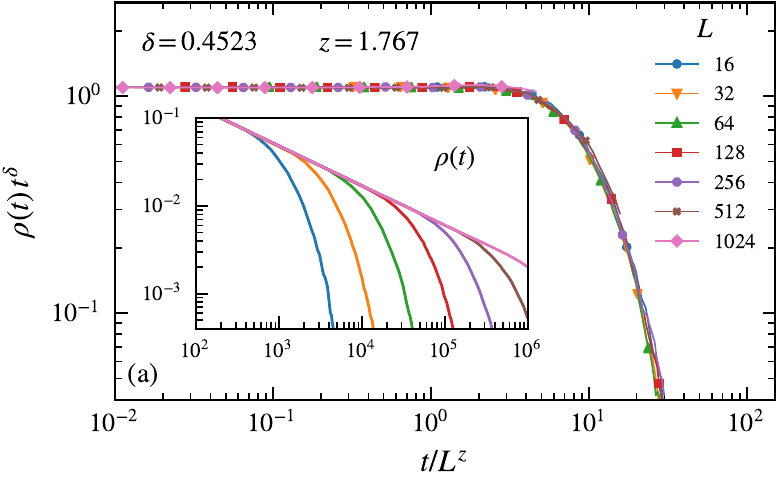}
	\includegraphics[width=\linewidth]{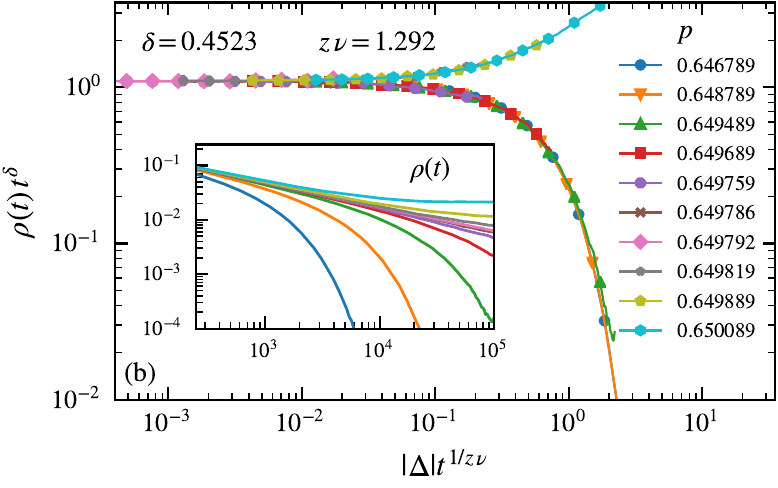}
	\caption{Scaling collapse plots for the CP. (a) Finite-size collapse of simulations starting from a fully occupied lattice at the critical point $p_c=0.649788$. (b) Data collapse in the off-critical region. The critical exponent estimates are given in the respective panels and the insets show the non-rescaled density as a function of time.}
	\label{fig:contact-results}		
\end{figure}

Once the critical probability is known, we perform decay simulations starting from a initially fully occupied lattice for different system sizes precisely at criticality and monitor the density $\rho(t)$ of active sites until the system reaches the absorbing state. This allows us to obtain the exponents $\delta$ and $z$ via data collapses according to
\begin{align}
\rho(L,t) = t^{-\delta} \tilde{\rho}\left(t/L^z\right),
\end{align}
where  $ \tilde{\rho} $ denotes a universal scaling function. In a second set of simulations we perform decay simulations for lattices of fixed size $L=1024$ in the vicinity of the critical point, which gives us the exponents $\delta$ and $\nu_\parallel = z\nu$ by means of the relation
\begin{align}
\rho(\Delta,t) = t^{-\delta} \hat{\rho}\left(\Delta t^{1/\nu_\parallel}\right),
\end{align}
where $\Delta=p-p_c$ is the distance from criticality and $\hat{\rho}$ a scaling function. Both scaling collapses, which turn out flawless, are shown in Fig.~\ref{fig:contact-results}, where we used the reference values, $\delta=0.4523(10)$, $z=1.7674(6)$ and $\nu_\parallel = 1.292(4)$ from Ref.~\cite{dickman1999}. In the top panel, all curves are averages over 1400 disorder realizations with 5 runs per realization, whereas in the bottom panel we used 250 disorder replicas and 5 runs per realization. The insets show the respective non-rescaled density as a function of time. This confirms the critical exponent values used and provides compelling evidence that CP on the random VG belongs to the clean DP universality class.

\section{Conclusion}
\label{sec:Conclusion}

Critical phenomena on Voronoi graphs have, to the best of our knowledge, not yet been investigated, focusing instead on lattices such as its dual, the Delaunay triangulation. In order to correct for this omission, we conducted large-scale numerical simulations of the classical Ising model and contact process on two-dimensional VGs constructed from randomly distributed sites. We establish reference values for the critical points of both models and, for the Ising model, also obtain the first correction-to-scaling exponent. Furthermore, using finite-size-scaling techniques we show that both systems display clean universal exponents at criticality, i.e., we reveal that the VG disorder is -- in the sense of the RG -- an irrelevant perturbation to their phase transitions.  Although we only analyzed two particular models, this result has implications for other classes of transitions as well. From the RG perspective, the correlation length exponent $\nu$ is directly related to the relevance of quenched disorder~\cite{harris1974,weinrib1983,kinzel1985,harris2016,barghathi2014}. For instance, the phase transition of regular (isotropic) percolation can be predicted to also remain unchanged on a Voronoi graph, since its exponent, $\nu=4/3$, is larger than that of both models considered. Moreover, our results are especially relevant for the search for a general disorder relevance criterion. The Voronoi graph has constant coordination number, similar to the constant coordination (CC) lattice we have recently introduced~\cite{schrauth2018a} and refined~\cite{schrauth2019a}. Studies of both the Ising and DP phase transition on the CC lattice~\cite{schrauth2018a,schrauth2019b} found disorder to be probably marginal in the Ising case and clearly \emph{relevant} for the DP universality class. The contrast of this result with the \emph{irrelevance} of the VG disorder shows that the absence of coordination number fluctuations is non-predictive of disorder relevance. Thus, coordination number fluctuations do not play a central role in determining the influence of quenched topological disorder on continuous phase transitions and a different direction, such as considering a measure of connectivity~\cite{schrauth2019b}, should be explored in the search for a general relevance criterion.

\begin{acknowledgments}
	We thank H.~Hinrichsen, F.~Goth and F.~P.~Toldin for helpful discussions. M.S.~thanks the Studienstiftung des deutschen Volkes for financial support. This work is part of the DFG research project Hi~744/9-1. The authors gratefully acknowledge the Gauss Centre for Supercomputing e.V. (www.gauss-centre.eu) for funding this project by providing computing time on the GCS Supercomputer SuperMUC at Leibniz Supercomputing Centre (www.lrz.de).
\end{acknowledgments}

%

\end{document}